\documentclass{jpsj3}
\usepackage{txfonts}
\usepackage{graphicx,amsmath,xspace}

\usepackage[usenames]{color}

\newcommand{\be}{\begin{equation}}
\newcommand{\ee}{\end{equation}}
\newcommand{\bea}{\begin{eqnarray}}
\newcommand{\eea}{\end{eqnarray}}
\newcommand{\alphaI}{{$\alpha$-(BEDT-TTF)$_2$I$_3$}\xspace}

\title{Algebraic structure of Dirac fermion state
  in $\alpha$-(BDET-TTF)$_2$I$_3$}

\author{
  \name{Takao \surname{Morinari}}$^1$
  \thanks{E-mail: morinari.takao.5s@kyoto-u.ac.jp}
  and
  \name{Yoshikazu \surname{Suzumura}}$^2$
}
\inst{
  $^1$Graduate School of Human and Environmental Studies, Kyoto University, Kyoto 606-8501, Japan\\
  $^2$Department of Physics, Nagoya University, Nagoya, 464-8602, Japan
}

\abst{
  We study the algebraic structure of the Dirac fermion state in 
  the organic conductor, \alphaI.
  We find a pair of generators
  for the Hamiltonian of \alphaI
  that describes the chirality of Dirac fermions.
  The phase parameters associated with those generators
  have an intimate and simple relationship with
  the positions of the Dirac points in the Brillouin zone.
  By making use of the form of the generators, a reduced form
  of the Hamiltonian is constructed
  that shares some characteristic features of the Dirac fermions 
  with \alphaI.
  For the reduced Hamiltonian we present
  the analytic expression for the Dirac point position
  that sits at {\it arbitrary} point in the Brillouin zone
  determined by the hopping parameters.
}

\kword{Dirac fermion, \alphaI, organic conductor}

\begin{document}


\maketitle

\section{Introduction}
Massless Dirac fermion spectrums in two-spatial dimension are now realized 
in various condensed matter systems.
A well-known system is graphene, a single atomic layer of graphite,
which is a two-dimensional hexagonal lattice of carbon atoms.
\cite{Novoselov2004}
The existence of the Dirac fermion spectrum was
clearly demonstrated by the observation 
of the half-integer quantum Hall effect.
\cite{Novoselov2005,Zhang2005}
Another example is a group of three-dimensional topological insulators
\cite{FuKaneMele07,MooreBalents07,Roy2009}
where the surface state exhibits a Dirac fermion spectrum.
A Dirac fermion spectrum appears even in superconductors.
In the antiferromagnetic state of iron pnictide superconductors,
a Dirac fermion spectrum\cite{Ran09,MorinariFeDirac2010} was observed in 
angle-resolved photoemission spectroscopy.\cite{Richard2010}
An organic compound, 
\alphaI [BEDT-TTF=bis(ethylenedithio)tetrathiafulvalene], 
which has a layered structure of BEDT-TTF molecules,
demonstrates that a Dirac fermion spectrum
appears by suppressing strong electron correlation effects.\cite{Katayama06}
Under ambient pressure, this compound undergoes a metal-insulator transition
due to a charge ordering induced by a strong electron correlation.
\cite{Kino1995,Seo2000,Takano2001,Wojciechowski2003}
This charge ordering is suppressed by applying pressure,
and transport measurements support the presence of 
a Dirac fermion spectrum\cite{Kajita1992,Tajima2000,Tajima06,Tajima2007} 
for pressures higher than 1.5GPa.
The presence of Dirac fermions is clearly demonstrated in the interlayer
magnetoresistance measurement \cite{Tajima09}
where the zero energy Landau level of Dirac fermions leads
to negative magnetoresistance.\cite{Osada08}

A natural question is how a Dirac fermion spectrum appears.
In case of graphene, the mechanism is trivial.
The Dirac fermion spectrum is realized as a result of symmetry
of the honeycomb lattice.
In case of topological insulators, essentially 
the spin-orbit interaction lead to a linear Dirac fermion energy 
spectrum.\cite{KaneMele2005a,KaneMele2005b,FuKaneMele07,MooreBalents07,Roy2009,HasanKane10}
In case of iron-pnictide superconductors, a characteristic
configuration of Fe and As leads to degeneracy of d-orbitals
at the $\Gamma$ point, and supports the appearance of 
Dirac fermion spectrum.\cite{Ran09}
Contrary to these systems, the mechanism of stabilizing 
the Dirac fermion spectrum in \alphaI  
is elusive.\cite{Mori2010,AsanoHotta2011}

In Dirac fermion systems the conduction and the valence bands touch 
at two inequivalent points in the Brillouin zone (BZ).
Usually reflecting symmetry of the system, the Dirac points appear
at high symmetric points or along symmetric lines in the BZ.
However, the Dirac points in \alphaI  are neither located 
at symmetric points of the BZ nor along symmetric lines in the BZ.
So the presence of the Dirac fermion spectrum of this system 
is called accidental degeneracy.\cite{Herring1937}
Furthermore, the positions of the Dirac points 
depend on the applied pressure.\cite{Katayama06,Kobayashi07}
The complicated four band structure of \alphaI  prevents 
revealing the essential
algebraic structure for the stabilization of the Dirac fermion spectrum.

In this paper, we study the underlying algebraic structure
for the Dirac fermion spectrum in \alphaI.
In spite of the complexity of the system,
{\it i.e.}, the presence of many hopping parameters 
and multi-orbital character,
there is a relatively simple algebraic structure that 
supports the Dirac fermion spectrum.
We show that two matrices $\Gamma_x$ and $\Gamma_y$ describe
chirality of Dirac fermions, and, interestingly,
phase parameters contained in these matrices
and the positions of the Dirac points
are intimately connected in a simple functional relationship.
Using the result, we construct a multi-orbital Hamiltonian with
a Dirac fermion spectrum whose Dirac points move in the BZ 
by changing the hopping parameters.

The paper is organized as follows.
In \S~\ref{sec_graphene} we consider graphene as a simple example
to illustrate the algebraic structure of a Dirac fermion system.
In \S~\ref{sec_model} we introduce the model for \alphaI,
and we show a simple relationship between
the matrices $\Gamma_x$ and $\Gamma_y$ and the positions of
the Dirac points.
In \S~\ref{sec_hamiltonian} we construct a multi-orbital Hamiltonian
that has a Dirac fermion spectrum whose Dirac points move
by changing the hopping parameters.
We also present the analytic expression for the Dirac point position
in the BZ.
In \S~\ref{sec_summary} we summarize the result.

\section{Algebraic structure of graphene}
\label{sec_graphene}
In order to illustrate the algebraic structure of Dirac fermions
we consider graphene as a simple example.
The purpose is to extract the algebraic structure
that is associated with chirality,
which is one of the most characteristic properties of Dirac fermions.

Usually chirality is defined by 
using eigenstates of Dirac fermions.\cite{CastroNeto09}
In case of two component Dirac fermion wave functions,
which is the most simplest Dirac fermion wave function,
one can define a chirality vector field by
$n_{k\alpha} =\langle \psi_k |\sigma_{\alpha}| \psi_k \rangle$
with $\psi_k$ being a two-component Dirac fermion wave function
at $k$ in the Brillouin zone
and $\sigma_{\alpha}$ being the Pauli matrices.
The chirality vector has either a vortex or anti-vortex configuration
around Dirac points.
  It is possible to construct this chirality vector field 
  directory from the Hamiltonian
  by introducing two matrices or generators.
  We illustrate how this construction is carried out in graphene.

The Hamiltonian for graphene is given by\cite{CastroNeto09}
\be
{\cal H} = \sum_{k} {\left( {\begin{array}{*{20}{c}}
{c_{A{k}}^\dag }&{c_{Bk}^\dag }
\end{array}} \right){H_k}\left( {\begin{array}{*{20}{c}}
{{c_{Ak}}}\\
{{c_{Bk}}}
\end{array}} \right)},
\label{eq_graphene}
\ee
where $c_{\alpha k}^{\dagger}$ ($c_{\alpha k}$) creates (annihilates)
an electron with the wave vector ${\bf k}$ on sublattice $\alpha = A,B$,
and
\be
H_k = \left( {\begin{array}{*{20}{c}}
0&{{\kappa _{{k}}}}\\
{{\kappa _{{k}}}^*}&0
\end{array}} \right),
\ee
with
\be
   {\kappa _{k}} =  
   - t\left[ {{e^{i{k_x}}} + {e^{i\left( { - \frac{1}{2}{k_x} 
             + \frac{{\sqrt 3 }}{2}{k_y}} \right)}} 
       + {e^{ - i\left( {\frac{1}{2}{k_x} + \frac{{\sqrt 3 }}{2}{k_y}} 
           \right)}}} \right].
\ee
We set the lattice constant unity.
Denoting a Dirac point in the BZ as ${\bf k}_D$, we find
$\kappa _{{\bf k}_D} = 0$.
Expanding $\kappa_{k}$ around ${\bf k}={\bf k}_D$, we find that
\[
{\left. {\frac{{\partial {\kappa _k}}}{{\partial {k_x}}}} 
\right|_{{\bf{k}} = {{\bf{k}}_D}}} 
= {C_x} \exp \left( { - \frac{i}{2}{k_{Dx}}} \right) i,
\hspace{2em}
{\left. {\frac{{\partial {\kappa _k}}}{{\partial {k_y}}}} 
\right|_{{\bf{k}} = {{\bf{k}}_D}}} 
= {C_y} \exp \left( { - \frac{i}{2}{k_{Dx}}} \right),
\]
where $C_x$ and $C_y$ are real constants.
Note that there is the factor 
$\exp \left( { - \frac{i}{2}{k_{Dx}}} \right)$ that depends 
on ${\bf k}_D$ in the right hand side of these equations.
We encounter similar factors for the case of \alphaI later.
Introducing the following matrix,
\[
{\Gamma _\alpha } = {\left. {\frac{{\partial 
{H_k}}}{{\partial {k_\alpha }}}} 
\right|_{{\bf{k}} = {{\bf{k}}_D}}},
\]
and representing these matrices as
$\Gamma_x = a_x (\sigma_x \cos\theta_x + \sigma_y \sin\theta_x)$
and
$\Gamma_y = a_y (\sigma_x \cos\theta_y + \sigma_y \sin\theta_y)$,
we find $|\theta_x - \theta_y|=\pi/2$.
  This phase difference $\pi/2$ comes from
  the phase difference between
  $\left. {{\partial {\kappa _k}}}/{{\partial {k_x}}}
  \right|_{{\bf{k}} = {{\bf{k}}_D}}$ and
  $\left. {{\partial {\kappa _k}}}/{{\partial {k_y}}}
  \right|_{{\bf{k}} = {{\bf{k}}_D}}$.
  If we define an inner-product between two matrices as
  $(A,B)={\rm tr}(AB)=\sum_{i,j}A_{ij}B_{ji}$,
  we find that 
  $(\Gamma_x,\Gamma_y) = a_x a_y \cos(\theta_x - \theta_y) = 0$.
  because of the phase difference $\pi/2$.

\begin{figure}
   \begin{center}
    \includegraphics[width=0.8 \linewidth]{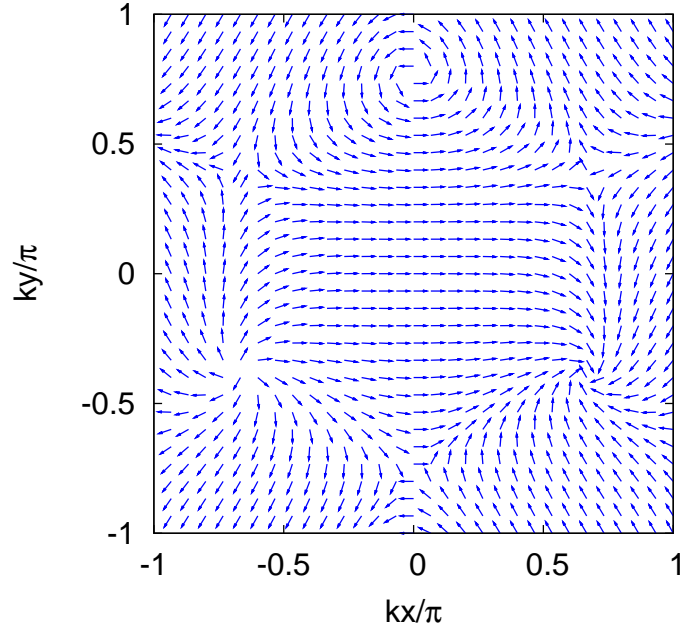}
   \end{center}
   \caption{ \label{fig_graphene}
     The vector field $(n_{kx},n_{ky})$ for graphene in {\bf k} space.
     The vector field is normalized to clarify the structure
     around the Dirac points.
     (We take the lattice constant as the unit of length.)
     The vector field forms vortex (anti-vortex)
     around K:$(2\pi/3)(1,1/\sqrt{3})$ (K$^\prime$:$(2\pi/3)(1,-1/\sqrt{3})$).
     The points K and K$^\prime$ are the Dirac points.
     (A similar result was presented in ref.~\citenum{SasakiWakabayashi2010}.)
    }
 \end{figure}

Now we construct the chirality vector directly from the Hamiltonian
using the matrices $\Gamma_{\alpha}$.
The $2 \times 2$ matrix $H_{k}$ has the following form 
around the Dirac point:
\be
H_{{\bf k}_D + \delta {\bf k}} = 
\Gamma_x \delta k_x + 
\Gamma_y \delta k_y + \cdots.
\ee
Suppose one moves around the Dirac point.
The path is represented by 
$\delta k_x = \delta k_0 \cos \phi$,
$\delta k_y = \delta k_0 \sin \phi$,
with $\phi$ changing from $0$ to $2\pi$.
We define a vector field $(n_{kx},n_{ky})$ as
\be
n_{kx} = {\rm tr}(\Gamma_x H_{\bf k}),\hspace{2em}
n_{ky} = {\rm tr}(\Gamma_y H_{\bf k}),
\ee
It is easy to see that $n_{kx}$ and $n_{ky}$ vanish at Dirac points.
As shown in Fig.~\ref{fig_graphene},
we find that this vector field rotates as
one goes around the Dirac point.
The vector field $(n_{kx},n_{ky})$ forms vortex (anti-vortex)
around the two inequivalent points.
These points correspond to the Dirac points in graphene.
Therefore, the vector field $(n_{kx},n_{ky})$
describes the chirality of the Dirac fermion.
Note that by symmetry the vector field $(n_{kx},n_{ky})$ can be rotated 
in the $k_x-k_y$ plane, or one can define
\be
n_{kx}' = {\rm tr}(\Gamma_x' H_{\bf k}),\hspace{2em}
n_{ky}' = {\rm tr}(\Gamma_y' H_{\bf k}),
\ee
with
$\Gamma_x' = \Gamma_x \cos \phi - \Gamma_y \sin \phi$
and
$\Gamma_y' = \Gamma_x \sin \phi + \Gamma_y \cos \phi$.
In Fig.~\ref{fig_graphene} we set $\Gamma_x = \sigma_x$
and $\Gamma_y = \sigma_y$ by making use of this symmetry.

In case of graphene, the analysis is trivial
because we have only four matrices,
the Pauli matrices $\sigma_x$, $\sigma_y$, $\sigma_z$,
and the unit matrix $\sigma_0$,
to construct $2 \times 2$ Hamiltonians.
Because of the absence of $\sigma_0$ and $\sigma_z$ components
in the graphene Hamiltonian, the space of matrices $\Gamma_{\alpha}$
is uniquely determined.

\section{Algebraic structure of \alphaI}
\label{sec_model}
Contrary to graphene, it is unclear how to find $\Gamma_{\alpha}$ 
matrices for the case of \alphaI because
the Hamiltonian is $4 \times 4$. (See, eq.~(\ref{eq_Hk}) below.)
There are 15 generators and one unit matrix
for constructing $4 \times 4$ Hermite matrix.
A suitably chosen set of those generators satisfy
commutation relations similar to that for 
$\sigma_x$, $\sigma_y$, and $\sigma_z$.
However, naive choices fail to describe chirality of 
Dirac fermions in \alphaI.

In order to find $\Gamma_{\alpha}$, we study 
the Hamiltonian of \alphaI.
We consider a single plane of BEDT-TTF molecules.
The unit cell consists of four BEDT-TTF molecules,
which are labeled by 1, 2, 3, and 4.
The Hamiltonian is given by\cite{Kobayashi04,Katayama06}
\be
   {\cal H} = \sum\limits_k {c_k^\dag {H_k}{c_k}}, 
\ee
where the spinor representation of the creation operator is 
$c_k^{\dagger} = \left( 
\begin{array}{cccc}
c_{1k}^{\dagger} &
c_{2k}^{\dagger} &
c_{3k}^{\dagger} &
c_{4k}^{\dagger} 
\end{array} \right)$ and 
\be
H_k = \left( {\begin{array}{*{20}{c}}
0&{{t_{c1}} + {t_{c2}}{e^{ - i{k_y}}}}&{{t_{p1}} 
- {t_{p4}}{e^{i{k_x}}}}&{{t_{p2}} - {t_{p3}}{e^{i{k_x}}}}\\
{{t_{c1}} + {t_{c2}}{e^{i{k_y}}}}&0&{{t_{p4}}{e^{i{k_y}}} - {t_{p1}}{e^{i\left( {{k_x} + {k_y}} \right)}}}&{{t_{p3}} - {t_{p2}}{e^{i{k_x}}}}\\
{{t_{p1}} - {t_{p4}}{e^{ - i{k_x}}}}&{{t_{p4}}{e^{ - i{k_y}}} - {t_{p1}}{e^{ - i\left( {{k_x} + {k_y}} \right)}}}&0&{{t_{c3}} + {t_{c4}}{e^{ - i{k_y}}}}\\
{{t_{p2}} - {t_{p3}}{e^{ - i{k_x}}}}&{{t_{p3}} - {t_{p2}}{e^{ - i{k_x}}}}&{{t_{c3}} + {t_{c4}}{e^{i{k_y}}}}&0
\end{array}} \right).
\label{eq_Hk}
\ee
Here hopping parameters $t_\eta$ ($\eta=c1,...,c4,p1,...,p4$) are 
assumed to be pressure dependent.
There are two types of pressure: uniaxial pressure and hydrostatic pressure.
For the purpose of investigating algebraic structure of eq.~(\ref{eq_Hk})
in wide hopping parameter ranges,
we consider a uniaxial pressure applied along the $a$-axis,
which is the stacking direction of BEDT-TTF molecules in the plane.
For the pressure dependence of $t_\eta = t_\eta (P_a)$, 
we take extrapolation formula given by\cite{Kobayashi04}
\be
t_\eta (P_a) = t_\eta (0) ( 1 + K_\eta P_a ).
\ee
The parameters $t_\eta (0)$ and $K_\eta$ are determined 
from the data at $P_a=0$ given in ref.~\citenum{Mori1984} 
and the data at $P_a=2$kbar given in ref.~\citenum{Kondo2005}.
The explicit values are\cite{Kobayashi04}
$t_{p1}(0)=0.140$, 
$t_{p2}(0)=0.123$, 
$t_{p3}(0)=-0.025$, 
$t_{p4}(0)=-0.062$, 
$t_{c1}(0)=0.048$, 
$t_{c2}(0)=-0.020$, 
$t_{c3}(0)=-0.028$, 
$t_{c4}(0)=-0.028$, 
$K_{p1}=0.011$,
$K_{p2}=0$,
$K_{p3}=0$,
$K_{p4}=0.032$,
$K_{c1}=0.167$,
$K_{c2}=-0.025$, and
$K_{c3}=K_{c4}=0.089$.

Our purpose is to find two matrices $\Gamma_x$ and $\Gamma_y$
that play a role similar to $\sigma_x$ and $\sigma_y$ in graphene.
Contrary to the Hamiltonian of graphene (\ref{eq_graphene}),
there are other orbitals.
Therefore, the situation is much more complicated.
In fact, one can find $\Gamma_x$ and $\Gamma_y$ from numerical calculations 
at a given pressure $P_a$.
However, it is hard to figure out the underlying algebraic structure.
Nevertheless we show that there is a simple algebraic structure
that governs the Dirac fermion spectrum of \alphaI.
We assume that $\Gamma_x$ and $\Gamma_y$ have the following form:
\be
  {\Gamma _x} = \left( {\begin{array}{*{20}{c}}
      {0}&{0}&{{e^{i{\phi _x}}}}&0\\
      {0}&{0}&0&{{e^{i{\theta _x}}}}\\
      {{e^{ - i{\phi _x}}}}&0&{0}&{0}\\
      0&{{e^{ - i{\theta _x}}}}&{0}&{0}
  \end{array}} \right), \hspace{2em}
  {\Gamma _y} = \left( 
  {\begin{array}{*{20}{c}}
      0&{{e^{i{\phi _y}}}}&{0}&{0}\\
      {{e^{ - i{\phi _y}}}}&0&{0}&{0}\\
      {0}&{0}&0&{{e^{i{\theta _y}}}}\\
      {0}&{0}&{{e^{ - i{\theta _y}}}}&0
  \end{array}} 
  \right).
  \label{eq_Gamma}
\ee
These forms are inferred from the forms of
${{\partial {H_{k}}}}/{{\partial {k_x}}}$ and
${{\partial {H_{k}}}}/{{\partial {k_y}}}$.
Some components are omitted so that ${\rm tr}(\Gamma_x\Gamma_y)=0$.
Furthermore, we take $\Gamma_x$ and $\Gamma_y$ so that 
$n_{{k}x}={\rm tr}(\Gamma_x H_k)$ depends on 
$t_{p1},t_{p2},t_{p3},t_{p4}$
and $n_{ky}={\rm tr}(\Gamma_y H_k)$ 
depends on $t_{c1},t_{c2},t_{c3},t_{c4}$.

\begin{figure}
   \begin{center}
    \includegraphics[width=0.8 \linewidth]{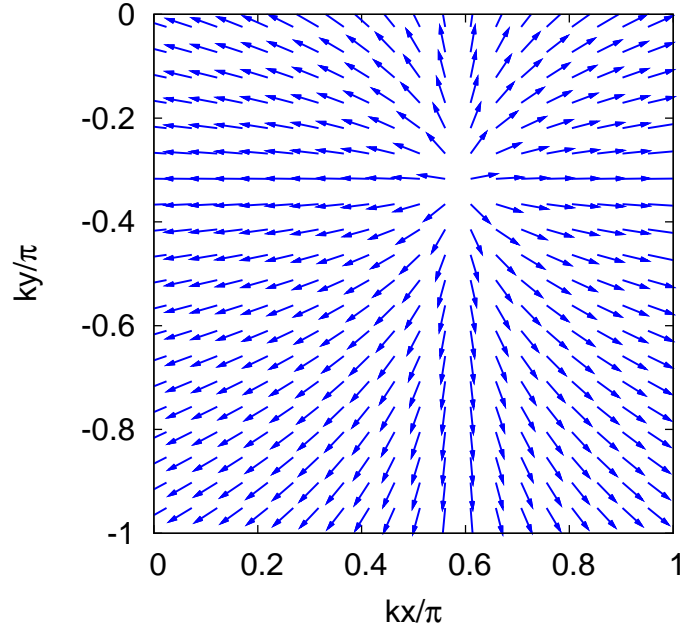}
   \end{center}
   \caption{ \label{fig_chirality_field}
     The vector field $(n_{kx},n_{ky})$ for \alphaI in {\bf k} space
       at $Pa=5$kbar.
     The vector field is normalized to clarify the structure
     around the Dirac point as in Fig.~\ref{fig_graphene}.     
     The phase parameters are $\theta_x = 161.85$ degree, 
     $\phi_y = 50.05$ degree.
    }
 \end{figure}

If we represent a Dirac point by $(k_{Dx},k_{Dy})$,
the phases in eq.~(\ref{eq_Gamma}) should depend on 
$k_{Dx}$ and $k_{Dy}$.
In order to make clear this dependence,
we shift $\theta_x$ and $\phi_y$
as $\theta_x \rightarrow \theta_x + k_{Dx}/2$
and
$\phi_y \rightarrow \phi_y - k_{Dy}/2$, respectively,
and fix $\phi_x$ and $\theta_y$ as
$\phi_x = \pi/2 + k_{Dx}/2$ and 
$\theta_y = \pi/2 - k_{Dy}/2$, respectively.
We find that the vector field $(n_{{k}x},n_{{k}y})$
has a vortex configuration in ${\bf k}$ space
as shown in Fig.~\ref{fig_chirality_field} on the plane
of $k_x$ and $k_y$.
The vortex moves in ${\bf k}$ space by changing 
the phase parameters $\phi_y$ and $\theta_x$.
The chirality of Dirac fermions in \alphaI is described by
$(n_{{k}x},n_{{k}y})$ by suitably choosing the phase parameters.

\begin{figure}
   \begin{center}
    \includegraphics[width=0.8 \linewidth]{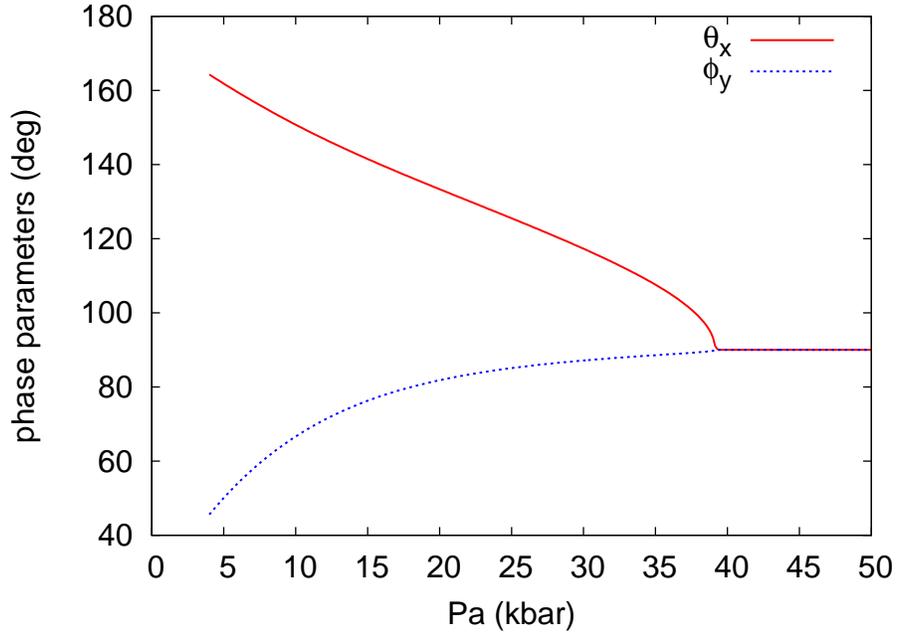}
   \end{center}
   \caption{ \label{fig_phase}
     Pressure dependence of the phase parameters $\theta_x$
     and $\phi_y$.
    }
 \end{figure}

\begin{figure}
   \begin{center}
    \includegraphics[width=0.8 \linewidth]{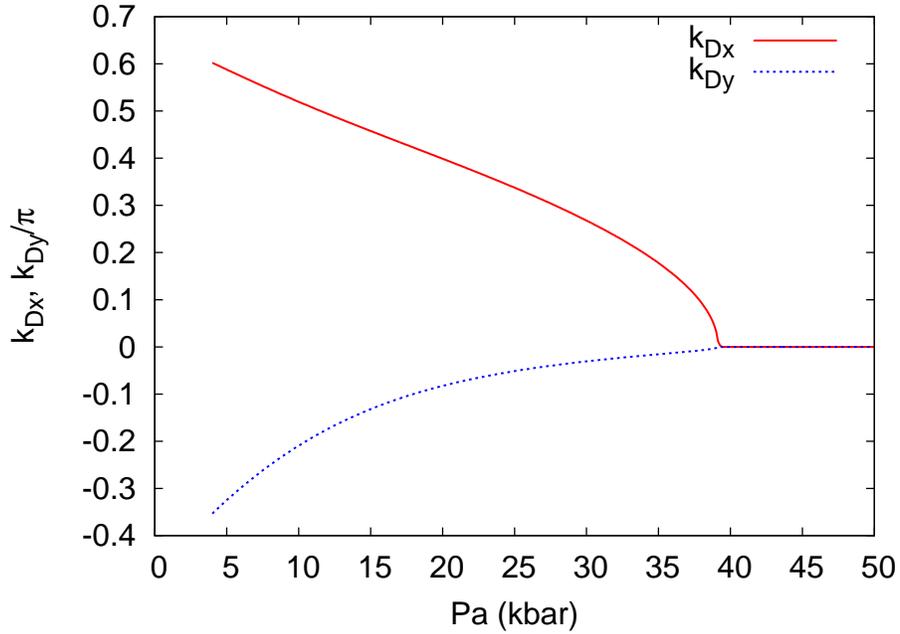}
   \end{center}
   \caption{ \label{fig_kD}
     Pressure dependence of the Dirac point in the third quadrant
     of BZ.
    }
 \end{figure}

\begin{figure}
   \begin{center}
    \includegraphics[width=0.8 \linewidth]{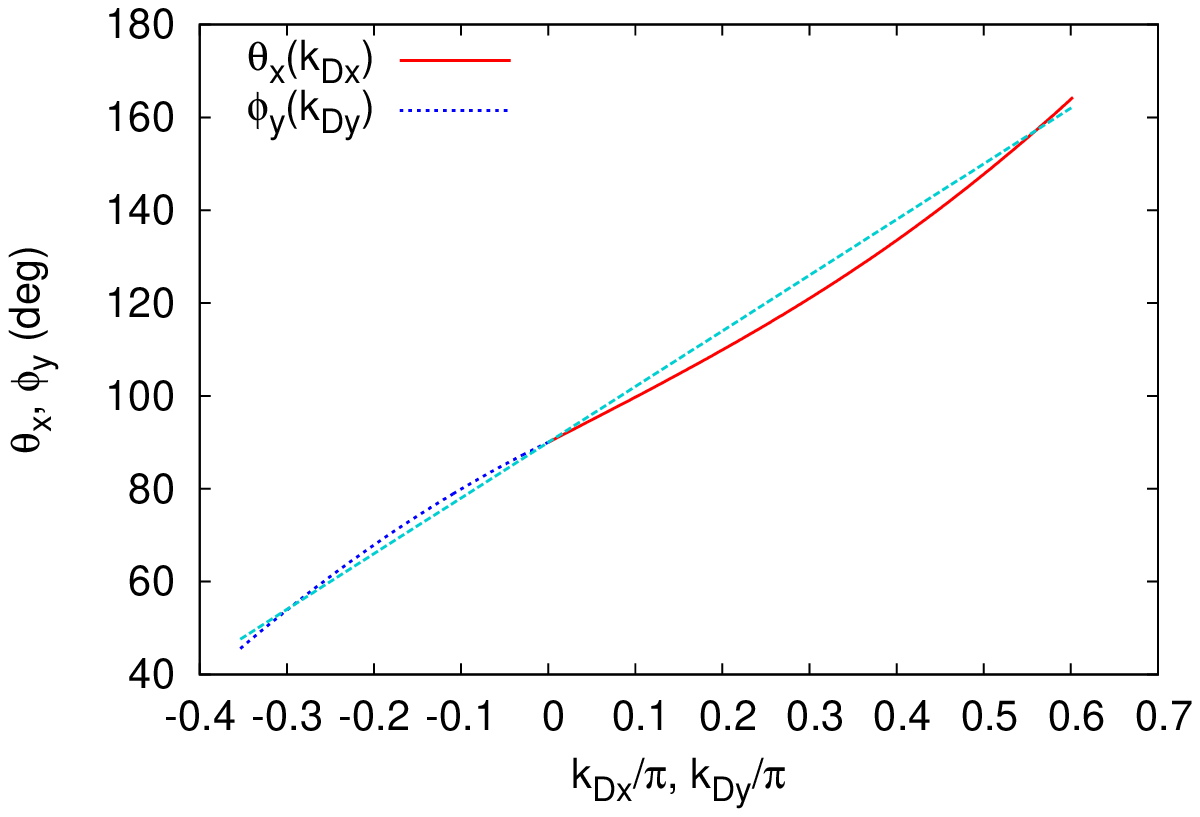}
   \end{center}
   \caption{ \label{fig_phase_kD}
     The phase parameters $\theta_x$ and $\phi_y$ as functions
     of $k_{Dx}$ and $k_{Dy}$, respectively.
     As shown by a dashed line, $\theta_x$ and $\phi_y$ 
     approximately linearly depend on $k_{Dx}$ and $k_{Dy}$, 
     respectively.
    }
 \end{figure}

The phase parameters are determined by solving
$(n_{{k}_D x},n_{{k}_D y}) = 0$:
\be
- {t_{p1}}\sin \frac{{{k_{Dx}}}}{2}
- {t_{p2}}\cos \left( {\theta _x} 
- \frac{{{k_{Dx}}}}{2} \right) 
+ {t_{p3}}\cos \left( {\theta _x} 
+ \frac{{{k_{Dx}}}}{2} \right) 
- {t_{p4}}\sin \frac{{{k_{Dx}}}}{2} = 0,
\ee

\be
{t_{c1}}\cos \left( {{\phi _y} - \frac{{{k_{Dy}}}}{2}} \right) + {t_{c2}}\cos \left( {{\phi _y} + \frac{{{k_{Dy}}}}{2}} \right) 
+ {t_{c3}}\sin {\frac{{{k_{Dy}}}}{2}} 
- {t_{c4}}\sin \frac{{{k_{Dy}}}}{2} = 0.
\ee
The pressure dependence of the phase parameters is shown
in Fig.~\ref{fig_phase}.
The Dirac points ${\bf k}_D$ are computed by diagonalizing
the Hamiltonian numerically.
Figure \ref{fig_kD} shows the pressure dependence 
of the positions of the Dirac points ${\bf k}_D$.
Note that the pressure dependence of $\theta_x$ 
is similar to that of $k_{Dx}$
and the pressure dependence of $\phi_y$ 
is similar to that of $k_{Dy}$.
Figure \ref{fig_phase_kD} shows $k_{Dx}$ dependence
of $\theta_x$ and $k_{Dy}$ dependence of $\phi_y$.
We find that the phase parameters are
approximately linearly dependent on the Dirac point coordinates,
that is,
\bea
\theta_x &=& \frac{2}{3} \left( k_{Dx} + \frac{3}{4}\pi \right),\nonumber \\
\phi_y &=& \frac{2}{3} \left( k_{Dy} + \frac{3}{4}\pi \right). \nonumber
\eea
This result demonstrates that underlying symmetry structure
is rather simple despite the fact that the Hamiltonian
has a complex form.
  The reason why we encounter the factor $2/3$ has not been clarified yet.
  The origin of this simple factor is left for a future study.

\section{Multiorbital Hamiltonian with Dirac fermion spectrum}
\label{sec_hamiltonian}
In the previous section, we have found matrices $\Gamma_x$ and $\Gamma_y$ 
that describe chirality of the Dirac fermions in \alphaI.
In this section, we construct a multi-orbital 
Dirac fermion Hamiltonian where chirality is 
{\it exactly} described by $\Gamma_x$ and $\Gamma_y$.
We show that the Hamiltonian has Dirac points whose positions
are hopping parameter dependent and 
not located at symmetric points in the BZ.
So characteristic features of Dirac fermions in \alphaI
are reproduced by this Hamiltonian.

\begin{figure}
   \begin{center}
    \includegraphics[width=0.8 \linewidth]{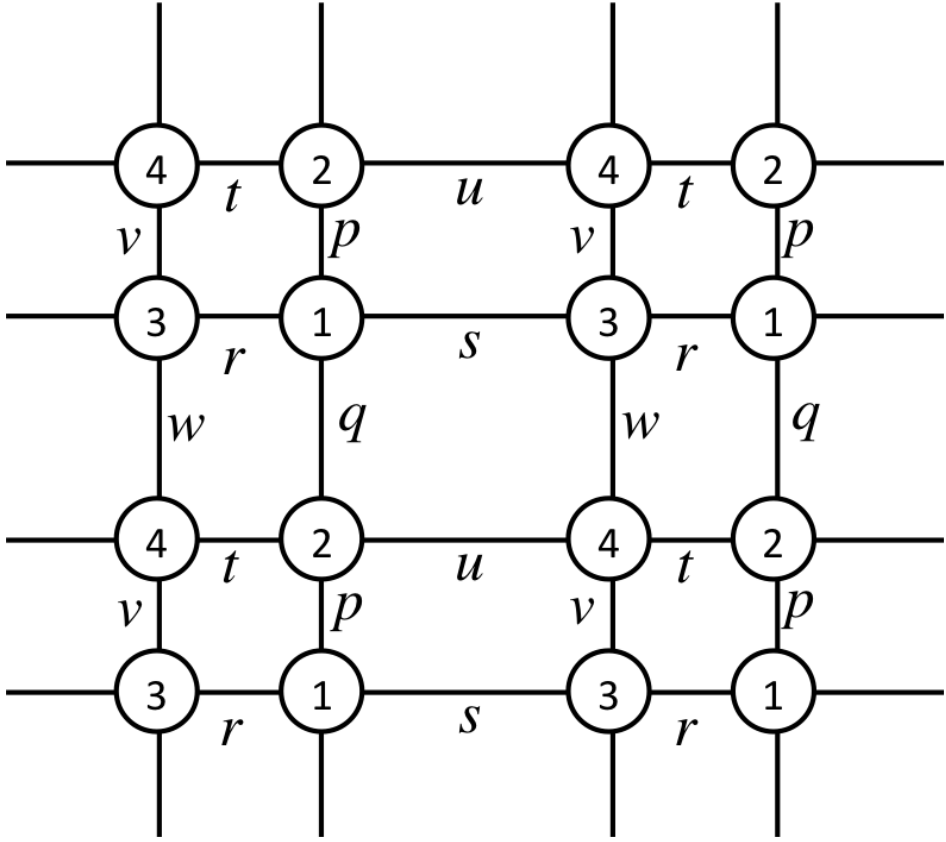}
   \end{center}
   \caption{ \label{fig_model}
     The hopping parameters of the reduced Hamiltonian eq.~(\ref{eq_Hred}).
     The unit cell contains four sites as in the case of
     \alphaI.
    }
 \end{figure}

From the form of eq.~(\ref{eq_Gamma}) and the fact that the chirality vector
field is obtained by taking trace of $\Gamma_{x} H_k$ and $\Gamma_{y} H_k$, 
we may consider the following Hamiltonian:
\be
   {H_{\bf{k}}} = \left( {\begin{array}{*{20}{c}}
       0&{p + q{e^{ - i{k_y}}}}&{r + s{e^{i{k_x}}}}&0\\
       {p + q{e^{i{k_y}}}}&0&0&{t + u{e^{i{k_x}}}}\\
       {r + s{e^{ - i{k_x}}}}&0&0&{v + w{e^{ - i{k_y}}}}\\
       0&{t + u{e^{ - i{k_x}}}}&{v + w{e^{i{k_y}}}}&0
   \end{array}} \right).
   \label{eq_Hred}
\ee
The hopping parameters,
$p$, $q$, $r$, $s$, $t$, $u$, $v$, and $w$,
which are real numbers,
are defined in Fig.~\ref{fig_model} in real space.
The Hamiltonian (\ref{eq_Hred}) is exactly diagonalized.
The eigenvalues are
\be
E_k^{\left( { \pm , \pm } \right)}
=  \pm \sqrt {A_k \pm \sqrt {{A_k^2} - {B_k^2}}}
\ee
where
\bea
A_k &=& \frac{1}{2}\left( {{{\left| {p + q{e^{ - i{k_y}}}} \right|}^2} 
+ {{\left| {r + s{e^{i{k_x}}}} \right|}^2} + {{\left| {t + u{e^{i{k_x}}}} 
\right|}^2} + {{\left| {v + w{e^{ - i{k_y}}}} \right|}^2}} \right), \\
B_k &=& \left| {\left( {p + q{e^{ - i{k_y}}}} \right)
\left( {v + w{e^{i{k_y}}}} \right) - \left( {r + s{e^{i{k_x}}}} \right)
\left( {t + u{e^{ - i{k_x}}}} \right)} \right|.
\eea

\begin{figure}
   \begin{center}
    \includegraphics[width=0.8 \linewidth]{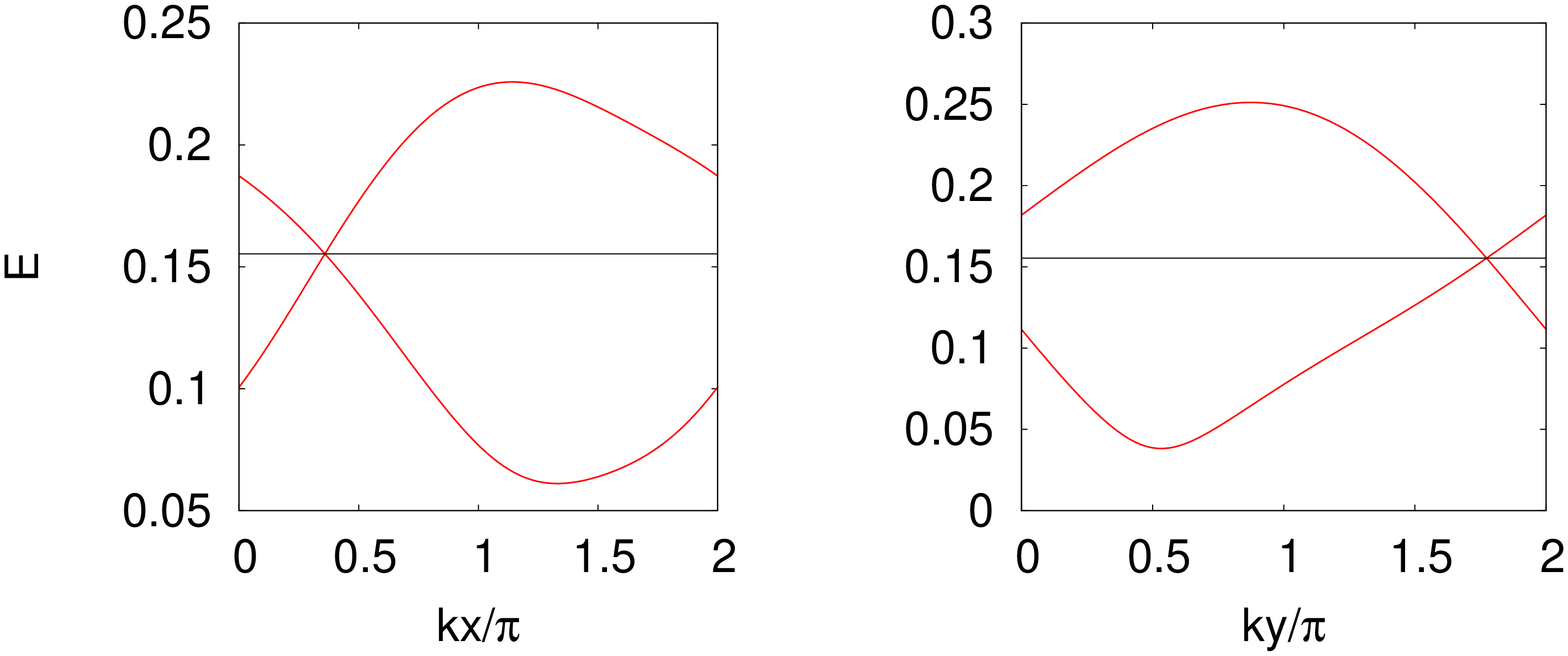}
   \end{center}
   \caption{ \label{fig_disp}
     The energy dispersions of the Hamiltonian (\ref{eq_Hred})
     along $k_x$-axis 
     with $k_y=1.773\pi$
     (left panel) 
     and $k_y$-axis 
     with $k_x=0.3587\pi$
     (right panel).
     The upper two bands, {\it i.e.}, $E_k^{(+,+)}$ and
     $E_k^{(+,-)}$ are shown.
     The Dirac point is at 
     $(k_x,k_y) = (0.3587, 1.773)\pi$.
     The horizontal solid line is the energy of the Dirac point.
     The parameter values are 
     $p = 0.08808$,
     $q = -0.0175$,
     $r = 0.1477$,
     $s = -0.0389$,
     $t = -0.025$,
     $u = -0.123$,
     $v = -0.04046$,
     $w = 0.1015$,
     and $\phi = 70$ degree.
     We set these values using the hopping parameter values
     of \alphaI at $P_a = 5$kbar except for $s$ and $w$.
     The parameters $s$ and $w$ are determined 
     from eqs.~(\ref{eq_constraint1}) and (\ref{eq_constraint2}).
    }
 \end{figure}

The upper two bands $E_k^{(+,+)}$ and $E_k^{(+,-)}$
touches at a point when $A_k=B_k$ is satisfied at a single ${\bf k}$ point.
Since there are many possibilities to satisfy this condition in general,
we limit ourselves to consider a simple case where
the equations
$v + w{e^{ - i{k_y}}} = {e^{i\phi }}\left( {p + q{e^{ - i{k_y}}}} \right)$
and 
$t + u{e^{i{k_x}}} =  - {e^{i\phi }}\left( {r + s{e^{i{k_x}}}} \right)$
are satisfied at some point $(k_x,k_y)$
with $\phi$ being a constant.
These equations are easily solved analytically
with respect to $k_x$ and $k_y$, and we find
\[
{k_x} = {\mathop{\rm Im}\nolimits} 
\left[ {\ln \left( { - \frac{{t + r{e^{i\phi }}}}{{u + s{e^{i\phi }}}}} 
\right)} \right],
\]
\[
{k_y} =  - {\mathop{\rm Im}\nolimits} 
\left[ {\ln \left( {\frac{{p{e^{i\phi }} - v}}{{w - q{e^{i\phi }}}}} 
\right)} \right].
\]
with constraints
$|pe^{i\phi}-v|=|w-qe^{i\phi}|$
and $|t+re^{i\phi}|=|u+se^{i\phi}|$.
The constraints are solved with respect to $w$ and $s$:
\be
w = q\cos \phi  \pm \sqrt {{p^2} + {v^2} 
- 2pv \cos \phi  - {q^2}{{\sin }^2}\phi },
\label{eq_constraint1}
\ee
\be
s = u\cos \phi  \pm \sqrt {{r^2} + {t^2}
+ 2r t\cos \phi  - {u^2}{{\sin }^2}\phi }.
\label{eq_constraint2}
\ee
We show an example in Fig.~\ref{fig_disp}
which always exhibits a zero gap state,
{\it i.e.}, the energy of the Dirac point 
does not intersect either $E_k^{(+,+)}$ or $E_k^{(+,-)}$ in the BZ
except at the Dirac points.
We clearly see that the energy dispersion is linear around
$(k_x,k_y) = (0.3587, 1.773)\pi$, which is the Dirac point.
Although we do not know the analytic solution for \alphaI
but we believe that the Hamiltonian (\ref{eq_Hred}) captures
essential symmetric properties of the Dirac fermions
of \alphaI.

\begin{figure}
   \begin{center}
    \includegraphics[width=0.8 \linewidth]{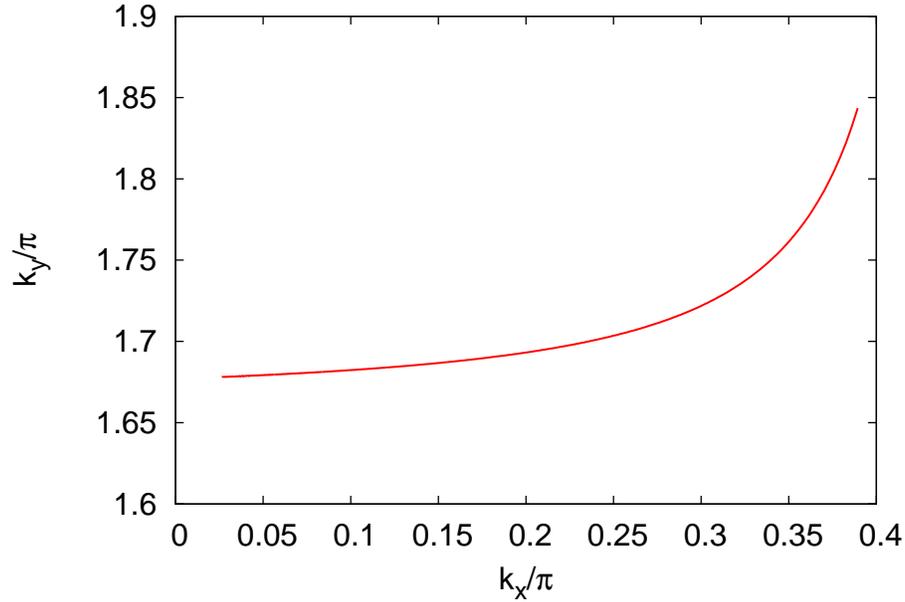}
   \end{center}
   \caption{ \label{fig_move}
     The motion of the Dirac point in the BZ.
     We take the hopping parameters as described in Fig.~\ref{fig_disp}
     with changing $P_a$.
    }
 \end{figure}

The fact that the Dirac point moves in the BZ 
by changing the hopping parameters is demonstrated
in Fig.~\ref{fig_move}.
Therefore, the Hamiltonian eq.~(\ref{eq_Hred})
shares the feature of the moving Dirac points
in the BZ with \alphaI.
We show one Dirac point ${\bf k}_D=(k_{Dx},k_{Dy})$
in Fig.~\ref{fig_move}.
The other Dirac point is at $(-k_{Dx},-k_{Dy})$ by symmetry.

It is interesting to point out a possible relationship between
the model and the $\pi$-flux state that was originally
proposed for the two-dimensional antiferromagnetic
Heisenberg model.\cite{AffleckMarston1988}
We shall discuss that \alphaI is characterized by
a $\pi$-flux state in a multi-orbital system in a future publication.

\section{Summary}
\label{sec_summary}
To summarize, we have investigated underlying algebraic structure
of the Dirac fermion spectrum of \alphaI.
We have found that chirality of Dirac fermions in \alphaI
is well described by the matrices $\Gamma_x$ and $\Gamma_y$
defined by eq.~(\ref{eq_Gamma}).
The phase factors in $\Gamma_x$ and $\Gamma_y$
have intimate relationships with the position of the Dirac point 
in the BZ,
and we have found a simple relationship between them.
A reduced form of the Hamiltonian has been constructed
from $\Gamma_x$ and $\Gamma_y$.
The Dirac point of the reduced Hamiltonian 
moves in the BZ by changing hopping parameters.
Although the analytic expression for the Dirac points of \alphaI 
is still unknown,
our reduced Hamiltonian has the analytic expression
for the Dirac points.
Therefore, the reduced Hamiltonian is a useful model to unveil
the mechanism of stabilizing the Dirac fermion spectrum in \alphaI.

\begin{acknowledgment}
The authors thank T.~Tohyama and 
A.~Kobayashi
for useful discussions.
This work was financially supported in part
by Grant-in-Aid for Special Coordination Funds for Promoting
Science and Technology (SCF), 
Scientific Research on Innovative Areas 20110002, 
and was also financially supported by 
a Grant-in-Aid for Special Coordination Funds
for Promoting Science and Technology (SCF) 
from the Ministry of Education, Culture, Sports, Science
and Technology in Japan,
a Grant-in-Aid for Scientific Research (A) 
on ``Dirac Electrons in Solids'' (No. 24244053) and 
a Grant-in-Aid for Scientific Research (C) (No. 24540370),
a Grant-in-Aid for Scientific Research (No. 23540403 and No. 23540403)
of The Ministry of Education, Culture, Sports, Science, 
and Technology, Japan.
\end{acknowledgment}


\bibliography{../../../references/tm_library2}

\end{document}